\documentclass{DISproc}

\newcommand{\be}{\begin{equation}}
\newcommand{\ee}{\end{equation}}
\newcommand{\bea}{\begin{eqnarray}}
\newcommand{\eea}{\end{eqnarray}}

\newcommand{\bfb}{\mbox{\boldmath $b$}}

\newcommand{\avkt}{\langle k_\perp^2 \rangle}
\newcommand{\avpt}{\langle p_\perp^2 \rangle}
\def\lsim{\mathrel{\rlap{\lower4pt\hbox{\hskip1pt$\sim$}}\raise1pt\hbox{$<$}}}
\def\gsim{\mathrel{\rlap{\lower4pt\hbox{\hskip1pt$\sim$}}\raise1pt\hbox{$>$}}}
\def\nostrocostruttino#1\over#2{\mathrel{\mathop{\kern 0pt \rlap
{\hbox{$#1$}}} \hbox{\kern-.135em $#2$}}}

\def\kt{k_\perp}
\def\bt{b_T}

\def\bbt{\bfb_T}

\def\pp{p_\perp}

\def\avp{\langle p_\perp ^2\rangle}

\begin{document}
%------------------------------------
\title{	Phenomenology of Sivers Effect \\ with TMD Evolution}

%for single authors the superscripts are optional
\author{{\slshape Mauro Anselmino$^1$, Mariaelena Boglione$^1$, Stefano Melis$^2$}\\[1ex]
$^1$ Dipartimento di Fisica Teorica, Universit\`a di Torino, \\ and INFN - Sezione di Torino,
             Via P.~Giuria 1, I-10125 Torino, Italy\\
$^2$ European Centre for Theoretical Studies in Nuclear Physics 
             and Related Areas (ECT*), \\
             Villa Tambosi, Strada delle Tabarelle 286, I-38123 Villazzano, 
             Trento, Italy}

% please enter the contribution ID for the DOI
\contribID{xy}

%\doi  % if there is an online version we will register DOIs

\maketitle

\begin{abstract}
Following the TMD evolution scheme recently proposed for the unpolarized 
and the Sivers distribution function, we propose a simple strategy to take 
into account this TMD $Q^2$ dependence in our phenomenological extraction 
of the Sivers function from SIDIS data. New results are presented and 
possible future applications are discussed. 
\end{abstract}

%\section{TMD evolution formalism}
The exploration of the 3-dimensional structure of the nucleon, both 
in momentum and in configuration space, is one of the major issues in 
high energy hadron physics. Information on this 3-dimensional structure 
is embedded in the Transverse Momentum Dependent distribution and 
fragmentation functions (TMDs). The Sivers function, which describes 
the number density of unpolarized quarks inside a transversely polarized 
proton, is particularly interesting, as it might provide information on the 
partonic orbital angular momentum.
%, as beautifully highlighted by W. Vogelsang, P. Mulders, F. Yuan and A. Bacchetta in their talks 
%(see their contributions in these proceedings). 

So far, all phenomenological fits have either neglected the QCD scale 
dependence of TMDs (which was unknown) or limited it to the collinear part 
of the unpolarized PDFs, according to the DGLAP evolution. Here, we present 
the first attempt to take into account the TMD evolution as proposed 
by Aybat, Collins, Qiu and Rogers~\cite{Collins:2011book, Aybat:2011zv,Aybat:2011ge} in the analysis of the Sivers asymmetry data and show how these 
new results compare with the previous extractions. Eventually, such a scheme 
will provide a complete TMD factorization framework for a consistent 
treatment of all SIDIS data. 

% A first application of the new TMD evolution equations to some limited samples of the HERMES 
% and COMPASS data~\cite{Aybat:2011ta} has indeed shown clear signs of  
% the $Q^2$ TMD evolution.     

In Ref.~\cite{Anselmino:2012aa} we showed how the QCD evolution equation 
of the TMDs in the coordinate space proposed in Refs. \cite{Aybat:2011zv} 
and \cite{Aybat:2011ge} can be expressed in a simplified way, taking
the renormalization scale $\mu^2$ and the regulating parameters $\zeta_F$ 
and $\zeta_D$ all equal to $Q^2$, as 
\be
\widetilde F(x, \bbt; Q) = \widetilde F(x, \bbt; Q_0)\> 
\widetilde R(Q, Q_0, \bt)\> \exp \left\{- g_K(b_T) \ln \frac{Q}{Q_0} \right\} 
\>, \label{Ftev}
\ee
where $\widetilde F$ can be either the unpolarized parton distribution,
$\widetilde F(x, \bbt; Q) = \widetilde f_{q/p}(x, \bbt; Q)$, the unpolarized 
fragmentation function $\widetilde F(x, \bbt; Q) = \widetilde D_{h/q}(z, 
\bbt; Q)$, or the first derivative, with respect to the parton impact 
parameter $b_T$, of the Sivers function, $\widetilde F(x, \bbt; Q) = 
\widetilde f_{1T}^{\prime \perp f}(x, \bbt; Q)$; 
$g_K(b_T)$ is an unknown, but universal
and scale independent, input function, 
while $\widetilde R(Q,Q_0,\bt)$ is the evolution kernel 
\be
\widetilde R(Q, Q_0, \bt)
\equiv 
\exp \left\{ \ln \frac{Q}{Q_0} \int_{Q_0}^{\mu_b} \frac{\rm d \mu'}{\mu'} \gamma_K(\mu') +
\int_{Q_0}^Q \frac{\rm d \mu}{\mu} 
\gamma_F \left( \mu, \frac{Q^2}{\mu^2} \right)\right\} 
\> \cdot \label{RQQ0}
\ee
The anomalous dimensions $\gamma_F$ and $\gamma_K$ appearing 
in Eq.~(\ref{RQQ0}), are given, at order 
${\cal O}(\alpha_s)$, by~\cite{Aybat:2011zv}
\be
\gamma_F(\mu; \frac{Q^2}{\mu^2}) = \alpha_s(\mu) \, \frac{C_F}{\pi}
\left( \frac{3}{2} - \ln \frac{Q^2}{\mu^2} \right)
\quad\quad\quad\quad 
\gamma_K(\mu) = \alpha_s(\mu) \, \frac{2 \, C_F}{\pi} \> \cdot
\label{gammas}
\ee
The $Q^2$ evolution is therefore driven by the functions $g_K(b_T)$ and 
$\widetilde R(Q,Q_0,\bt)$. While the latter, Eq.~(\ref{RQQ0}), can be easily evaluated, 
numerically or even analytically, the former, is essentially 
unknown and will need to be taken from independent experimental inputs.

The appropriate Fourier transforms allow us to obtain the distribution 
and fragmentation functions in the momentum space:
\begin{align}
\widehat f_{q/p}(x, \kt; Q) &= \frac{1}{2\pi} \int_0^\infty \!\!\!{\rm d} b_T 
\> b_T \> J_0(k_\perp b_T) \> \widetilde f_{q/p}(x, b_T; Q) \>,
\label{TMDunpf}
\\
\widehat D_{h/q}(z, \pp; Q) &= \frac{1}{2\pi} \int_0^\infty \!\!\!{\rm d} b_T 
\> b_T \> J_0({\rm k}_T b_T) \> \widetilde D_{h/q}(z, \bt; Q) \>,
\label{TMDunpD}
\\
\widehat f_{1T}^{\perp f}(x, k_\perp; Q) &= \frac{-1}{2\pi k_\perp} \int_0^\infty 
\!\!\! {\rm d} b_T \> b_T \> J_1(k_\perp b_T) \> 
\widetilde f_{1T}^{\prime \,\perp q}(x, b_T; Q) \>, \label{TMDsiv}
\end{align}
where $J_0$ and $J_1$ are Bessel functions, while $\widehat f_{q/p}$ is 
the unpolarized TMD distribution function for a parton of flavor $q$ inside 
a proton, $\widehat D_{h/q}$ is the unpolarized TMD fragmentation function 
for hadron $h$ inside a parton $q$ and $\widehat f_{1T}^{\perp q}$ is the 
Sivers distribution describing unpolarized partons inside a transversely 
polarized proton.
% , as:
% %
% \bea
% \widehat f_{q/p^\uparrow}(x, \bfk_\perp, \bfS; Q) &=& 
% \widehat f_{q/p}(x, k_\perp; Q) - \widehat f_{1T}^{\perp q}(x, k_\perp; Q)\frac{\epsilon_{ij} \, k_\perp^i \, 
% S^j}{M_p} \label{Siv1} \\
% &=& \widehat f_{q/p}(x, k_\perp; Q)
% + \frac 12 \Delta^N \widehat f_{q/p^\uparrow}(x, k_\perp; Q)\frac{\epsilon_{ij} 
% \, k_\perp^i \, S^j}{k_\perp} \> \cdot \label{Siv2}
% \eea

%\section{Parameterization of unknown functions}

The unknown input functions $g_K(\bt)$ and $\widetilde F(x, b_T; Q_0)$ inside 
Eq.~(\ref{Ftev}) have to be appropriately parameterized. 
As already anticipated, $g_K(\bt)$ is a non-perturbative, but universal function, 
which in the literature is usually parameterized in a quadratic form: $g_K(b_T) = \frac12 \, g_2 \, b_T^2$.  
As in 
Ref.~\cite{Aybat:2011ge} %and \cite{Aybat:2011ta}, 
we will adopt the results 
provided by a recent fit of Drell-Yan data~\cite{Landry:2002ix}, and assume $g_2 = 0.68$ GeV$^2$.
%
% \be
% g_K(b_T) = \frac12 \, g_2 \, b_T^2 
% \>.
% \label{gk}
% \ee
The input functions $\widetilde F(x, b_T; Q_0)$ are parameterized by requiring that their 
Fourier-transforms, which give the corresponding TMD functions in the 
transverse momentum space, coincide with the previously adopted 
$\kt$-Gaussian forms, with the $x$ dependence factorized out. 
As shown in Ref.~\cite{Anselmino:2012aa}, one finds
\begin{align}
\widetilde f_{q/p}(x, \bt; Q)& =  f_{q/p}(x,Q_0) \;\widetilde R(Q, Q_0, \bt)\;
\exp \left\{-b_T^2 \left(\alpha ^2\,  + \frac{g_2}{2} \ln \frac{Q}{Q_0}\right) \right\} 
\label{evF-f}
\\
\widetilde D_{h/q}(z, \bt; Q) &= \frac{1}{z^2} D_{h/q}(z,Q_0) \;
\widetilde R(Q, Q_0, \bt)\;
\exp \left\{- b_T^2 \left(\beta ^2\,  + \frac{g_2}{2} 
\ln \frac{Q}{Q_0}\right) \right\} 
\label{evF-D}
\\
\widetilde f_{1T}^{\prime \perp }(x, \bt; Q)\;\; &=  -2 \, \gamma^2 \,
f_{1T}^{\perp}(x; Q_0) \, \widetilde R(Q,Q_0,b_T) \, \bt \, 
\exp \left\{-b_T^2 \left( \gamma^2\, + \frac{g_2}{2} \ln \frac{Q}{Q_0} 
\right) \right\} \,
\label{evF-Sivers}
\end{align}
with $\alpha ^2=\avkt/4$, $\beta^2 = \avpt/(4z^2)$, $4 \, \gamma^2 \equiv \avkt _S = \frac{M_1^2 \, \avkt}{M_1^2 +\avkt}$, 
%$g_2$ given in Eq.~(\ref{gk}) 
and $\widetilde R(Q, Q_0, \bt)$ in Eq.~(\ref{RQQ0}).
%\be
%4 \, \gamma^2 \equiv \avkt _S = \frac{M_1^2 \, \avkt}{M_1^2 +\avkt}
%\label{gamma}
%\ee 

Eqs.~(\ref{evF-f})-(\ref{evF-Sivers}) show that the $Q^2$ evolution is 
controlled by the logarithmic $Q$ dependence of the $b_T$ Gaussian width, 
together with the factor $\widetilde R(Q, Q_0, \bt)$: for increasing values 
of $Q^2$, they are responsible for the typical broadening effect already 
observed in Refs.~\cite{Aybat:2011zv} and~\cite{Aybat:2011ge}.

%\section{Analytical solution of the TMD Evolution Equations}

As $R(Q,Q_0,b_T)$ shows a weak dependence on (large) $b_T$ (i.e. 
small $\kt$), we can assume $R(Q,Q_0,b_T)$ to be constant in $b_T$
and compute the Fourier transforms of the evolution equations~\eqref{evF-f},~\eqref{evF-D} and \eqref{evF-Sivers} analytically, to find
\begin{align}
\widehat f_{q/p}(x,\kt;Q)&=f_{q/p}(x,Q_0)\; R(Q,Q_0) \; 
\frac{ e^{-\kt ^2/w^2}}{\pi\,w^2} \label{unp-gauss-evol} \\
%\ee
%
%\be
\widehat D_{h/q}(z,\pp;Q) &=  D_{h/q}(z,Q_0)\; R(Q,Q_0) \; 
\frac{e^{-\pp^2/w^2_{\!F}}}{\pi w^2 _{\!F}} 
\label{D-gauss-evol} \\
%\ee
%
%\be
\Delta^N \widehat f_{q/p^\uparrow}(x,\kt;Q)&=\frac{\kt}{M_1}\,\sqrt{2 e}\,
\frac{\avkt _S^2}{\avkt}\,\Delta^N f_{q/p^\uparrow}(x,Q_0)\,R(Q,Q_0)\, 
\frac{e^{-\kt^2/w_S^2}}{\pi w_S^4} \>, \label{siv-gauss-evol}
\end{align}
%\ee
%
where $f_{q/p}(x,Q_0)$ and $D_{h/q}(z,Q_0)$  are the usual integrated PDF 
evaluated at the initial scale $Q_0$, and $\Delta^N f_{q/p^\uparrow}(x,Q_0)$ 
gives the $x$ dependence of the Sivers function \cite{Anselmino:2012aa}
[$- (2 k_\perp/M_p) \widehat f_{1T}^\perp = \Delta^N \widehat 
f_{q/p^\uparrow}$]. Most importantly, $w^2$, $w^2 _{\!F}$ and $w_S^2$ 
are the ``evolving'' Gaussian widths, defined as:
\be
w^2  =\avkt + 2\,g_2 \ln \frac{Q}{Q_0}\>  \,,\quad \quad
w _{\!F}^2   =\avp + 2 z^2 g_2 \ln \frac{Q}{Q_0} \> \,,\quad \quad
w^2_S  =\avkt _S + 2 g_2 \ln \frac{Q}{Q_0}\> \cdot \label{w}
\ee
Notice that the $Q^2$ evolution of the TMD PDFs is now 
determined by the overall factor $R(Q,Q_0)$ and, most crucially, by the 
$Q^2$ dependent Gaussian width $w(Q,Q_0)$. 
%
%%%%%%%%%%%%%%%%%%%%%%%%%%%%%%%%%%%%%%%%%%%%%%%%%%%%%%%%%%%%%%%%%%%%%%%%%%%%%%%
\begin{table}[t]
\caption{$\chi^2$ contributions corresponding to our three fits, for some experimental data sets of HERMES and 
COMPASS experiments.\label{tab:chi-sq}}
\begin{tabular}{cccc}
\toprule
    & TMD Evolution (exact)   & TMD Evolution (analyt.) & DGLAP Evolution\\
\noalign{\vspace{2pt}}
\cline{2-4}
\noalign{\vspace{2pt}}
    & $\chi^2_{tot}\;=255.8$  & $\chi^2_{tot}\;=275.7$   &   $\chi^2_{tot}\;=315.6$\\
\noalign{\vspace{1pt}}
    & $\chi^2_{d.o.f}=\;1.02$ & $\chi^2_{d.o.f}=\;1.10$  & $\chi^2_{d.o.f}=\;1.26$\\
\midrule
                          & $\chi^2_x\;=10.7$      &  $\chi^2_x\;=12.9$      & $\chi^2_x\;=27.5$   \\
HERMES $\pi^+$            & $\chi^2_z\;=\;4.3$     &  $\chi^2_z\;=\;4.3$     & $\chi^2_z\;=8.6$     \\
                          & $\chi^2_{P_T}\!=9.1$   &  $\chi^2_{P_T}\!=10.5$  & $\chi^2_{P_T}\!=22.5$ \\
\midrule
                          & $\chi^2_x=\;6.7$       &  $\chi^2_x=11.2$        & $\chi^2_x=29.2$   \\
 COMPASS $h^+$            & $\chi^2_z=17.8$        &  $\chi^2_z=18.5$        & $\chi^2_z=16.6$   \\
                          & $\chi^2_{P_T}\!=12.4$  &  $\chi^2_{P_T}\!=24.2$  &  $\chi^2_{P_T}\!=11.8$   \\
\bottomrule
\end{tabular}
\end{table}
%%%%%%%%%%%%%%%%%%%%%%%%%%%%%%%%%%%%%%%%%%%%%%%%%%%%%%%%%%%%%%%%%%%%%%%%%%%%%%
%

It is interesting to point that the evolution factor $R(Q,Q_0)$, 
controlling the TMD evolution, is the same for all functions 
(TMD PDFs, TMD FFs and Sivers ) and is flavor independent: consequently it 
will appear, squared, in both numerator and denominator of the Sivers azimuthal 
asymmetry and, approximately, cancel out. Therefore, we can safely conclude 
that most of the TMD evolution of azimuthal asymmetries is controlled by the 
logarithmic $Q$ dependence of the $\kt$ Gaussian widths $w^2(Q,Q_0)$,
Eq.~(\ref{w}). 

The aim of our paper is to analyze the available polarized SIDIS data from 
the HERMES and COMPASS collaborations in order to understand whether or not 
they show signs of the TMD evolution proposed in Ref.~\cite{Aybat:2011ge}. 
In particular we perform three different data fits of the SIDIS Sivers 
single spin asymmetry $A^{\sin(\phi_h-\phi_S)}_{UT}$ measured by HERMES and 
COMPASS: a fit (TMD-fit) in which we adopt the TMD evolution equations 
of Eqs.~(\ref{evF-f})-(\ref{evF-Sivers});% and (\ref{TMDunpf})-(\ref{TMDsiv});
a second fit (TMD-analytical-fit) in which we apply the same TMD evolution, 
but using the analytical approximation of
Eqs.~(\ref{unp-gauss-evol}), (\ref{D-gauss-evol}) and (\ref{siv-gauss-evol});
a fit (DGLAP-fit) in which we follow our previous work, as done so far 
in Ref.~\cite{Anselmino:2008sga,Anselmino:2011gs}, using the DGLAP evolution 
equation only in the collinear part of the TMDs. 
%
%%%%%%%%%%%%%%%%%%%%%%%%%%%%%%%%%%%%%%%%%%%%%%%%%%%%%%%%%%%%%%%%%%%%%%%%%%%%%%%%
\begin{wrapfigure}{r}{0.5\textwidth}
\hspace*{-0.9cm}
    \includegraphics[width=0.25\textwidth,angle=-90]{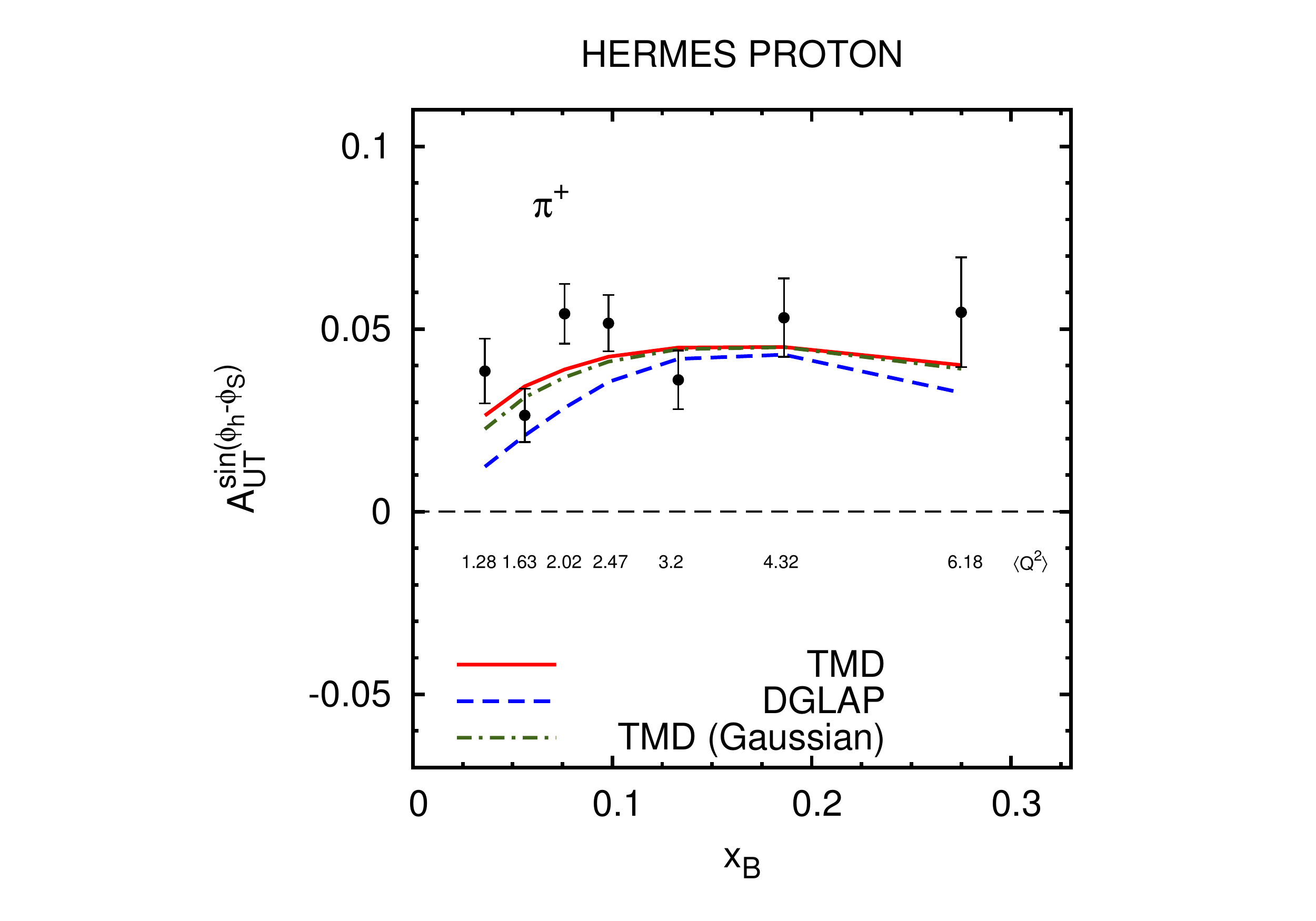}\hspace*{-1.5cm}
   \includegraphics[width=0.25\textwidth,angle=-90]{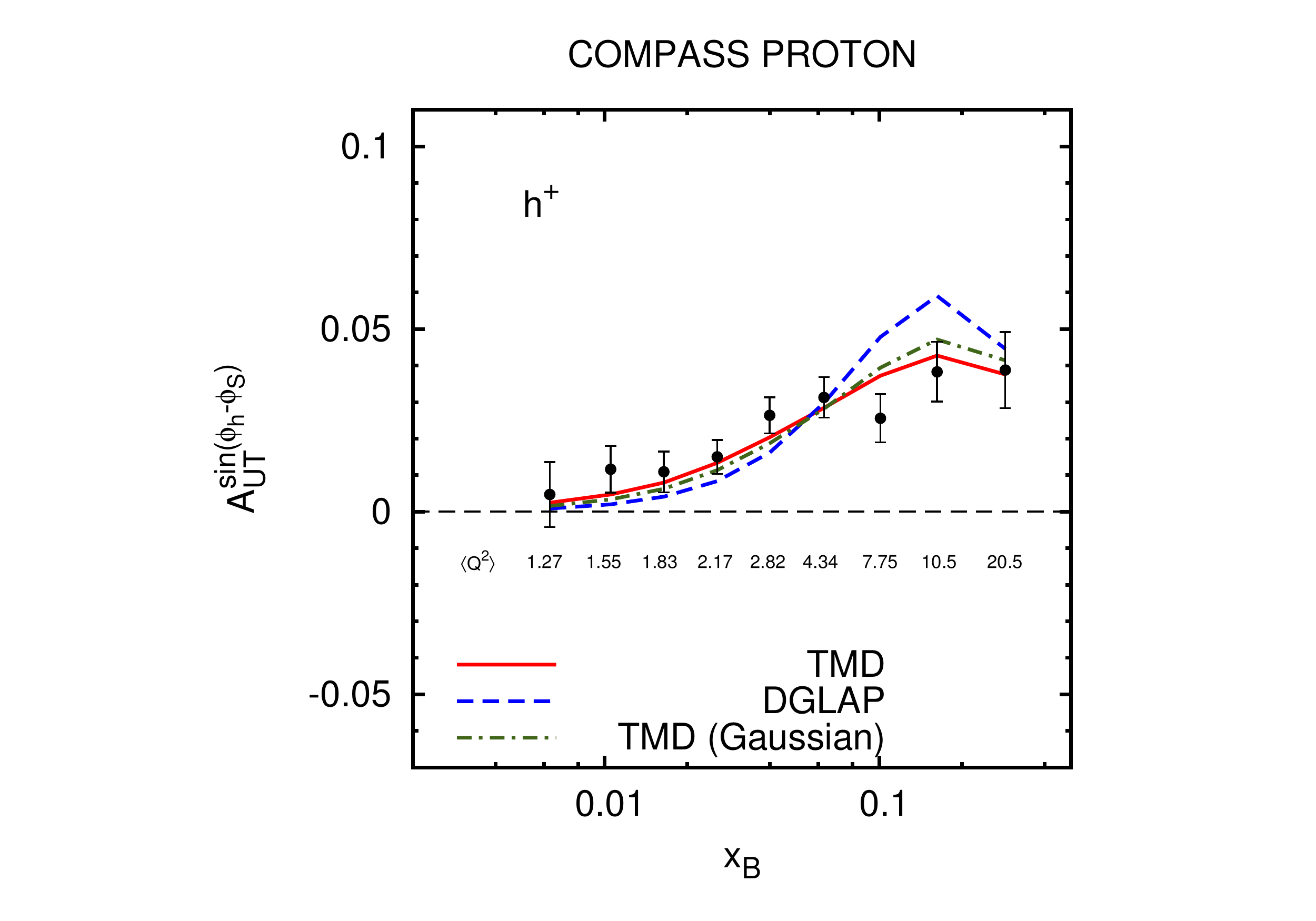}
  \caption{\label{fig:hermes-compass-confronto}The results obtained from our fit of the SIDIS $A_{UT}^{\sin{(\phi_h-\phi_S)}}$ 
Sivers asymmetries applying TMD evolution (red, solid lines) are compared with 
the analogous results found by using DGLAP evolution equations (blue, dashed lines). 
The green, dash-dotted lines correspond to the results obtained by 
using the approximated analytical TMD evolution. 
The experimental data are from HERMES~\cite{:2009ti} (left panel) and 
COMPASS~\cite{Bradamante:2011xu} (right panel) Collaborations. }
\end{wrapfigure}
%%%%%%%%%%%%%%%%%%%%%%%%%%%%%%%%%%%%%%%%%%%%%%%%%%%%%%%%%%%%%%%%%%%%%%%%%%%%%%%%
Table~I shows the main results of our fitting procedure. 
The best total $\chi^2_{tot}$, 
which amounts to $256$, is obtained by using the TMD evolution, followed 
by a slightly higher $\chi^2_{tot}$ of the analytical approximation, and 
a definitely larger $\chi^2_{tot} \simeq 316$ corresponding to the DGLAP fit.
The difference 
of about 60 $\chi^2$-points between the TMD and the DGLAP fits 
is heavily concentrated in the asymmetry for 
$\pi^+$ production at HERMES and for $h^+$ production at COMPASS, especially when this asymmetry is observed as a function 
of the $x$-variable.
It is important to stress that, as $x$ is directly proportional to $Q^2$ 
through the kinematical relation $Q^2=x\,y\,s$, the $x$ behavior of the 
asymmetries is intimately connected to their $Q^2$ evolution. While the HERMES 
experimental bins cover a very modest range of $Q^2$ values, from $1.3$ GeV$^2$ 
to $6.2$ GeV$^2$, COMPASS data raise to a maximum $Q^2$ of $20.5$ GeV$^2$, 
enabling to test more severely the TMD $Q^2$ evolution in SIDIS.   
These aspects are illustrated in Fig.~\ref{fig:hermes-compass-confronto}, 
where the SIDIS Sivers asymmetries  $A^{\sin(\phi_h-\phi_S)}_{UT}$ obtained 
in the three fits are shown in the same plot. It is evident that the DGLAP 
evolution seems to be unable to describe the correct $x$ trend, {\it i.e.} 
the right $Q^2$ behavior, while the TMD evolution (red solid line) follows 
much better the large $Q^2$ data points, corresponding to the last $x$-bins 
measured by COMPASS.

In conclusions, 
we have analyzed the Sivers effect by up-grading old fits with the
addition of the most recent HERMES and COMPASS SIDIS data, and by applying, 
for the first time, TMD evolution equations. We have compared the results 
obtained using TMD evolution equations with those found by considering 
only the DGLAP evolution of the collinear part of the TMDs. Our results 
give evidence that SIDIS data support the TMD evolution scheme, although
further experimental data, covering a wider range of $Q^2$ values, are 
necessary to confirm this.

{\raggedright
\begin{footnotesize}
% IF YOU DO NOT USE BIBTEX, USE THE FOLLOWING SAMPLE SCHEME FOR THE REFERENCES
% ----------------------------------------------------------------------------

% ----------------------------------------------------------------------------

% IF YOU USE BIBTEX,
% - DELETE THE TEXT BETWEEN THE TWO ABOVE DASHED LINES
% - UNCOMMENT THE NEXT TWO LINES AND REPLACE 'smith_joe.bib' WITH YOUR
%   FILE(S)

% \bibliographystyle{DISproc}
% \bibliography{smith_joe.bib}

\begin{thebibliography}{99}


\bibitem{Collins:2011book}
  J.~C.~Collins, Foundations of Perturbative QCD, Cambridge Monographs 
  on Particle Physics, Nuclear Physics and Cosmology, No. 32, Cambridge 
  University Press, Cambridge, 2011.

\bibitem{Aybat:2011zv} 
  S.~M.~Aybat and T.~C.~Rogers,
  %``TMD Parton Distribution and Fragmentation Functions with QCD Evolution,''
  Phys.\ Rev.\ D {\bf 83}, 114042 (2011)
  [arXiv:1101.5057 [hep-ph]].
   
\bibitem{Aybat:2011ge}
  S.~M.~Aybat, J.~C.~Collins, J.~-W.~Qiu and T.~C.~Rogers,
  %``The QCD Evolution of the Sivers Function,''
  arXiv:1110.6428 [hep-ph].
  
\bibitem{Anselmino:2012aa} 
  M.~Anselmino, M.~Boglione and S.~Melis,
  %``A Strategy towards the extraction of the Sivers function with TMD evolution,''
  arXiv:1204.1239 [hep-ph].
  %%CITATION = ARXIV:1204.1239;%%
  
\bibitem{Landry:2002ix} 
  F.~Landry, R.~Brock, P.~M.~Nadolsky and C.~P.~Yuan,
  %``Tevatron Run-1 $Z$ boson data and Collins-Soper-Sterman resummation 
  %formalism,''
  Phys.\ Rev.\ D {\bf 67}, 073016 (2003)
  [hep-ph/0212159].
  
\bibitem{Anselmino:2008sga} 
  M.~Anselmino, M.~Boglione, U.~D'Alesio, A.~Kotzinian, S.~Melis, F.~Murgia,   
  A.~Prokudin and C.~T\"urk,
  Eur.\ Phys.\ J.\ A {\bf 39}, 89 (2009)
  [arXiv:0805.2677 [hep-ph]]. 
  
  \bibitem{Anselmino:2011gs} 
  M.~Anselmino, M.~Boglione, U.~D'Alesio, S.~Melis, F.~Murgia and A.~Prokudin,
  %``Sivers distribution functions and the latest SIDIS data,''
  arXiv:1107.4446 [hep-ph].

  \bibitem{:2009ti}
  A.~Airapetian {\it et al.}  [HERMES Collaboration],
  %``Observation of the Naive-T-odd Sivers Effect in Deep-Inelastic
  %Scattering,''
  Phys.\ Rev.\ Lett.\  {\bf 103}, 152002 (2009)
  [arXiv:0906.3918 [hep-ex]].

\bibitem{Bradamante:2011xu} 
  F.~Bradamante [COMPASS Collaboration],
  %``New COMPASS results on Collins and Sivers asymmetries,''
  arXiv:1111.0869 [hep-ex].
  %%CITATION = ARXIV:1111.0869;%%
  
  
  
% \bibitem{Collins:1981va} 
%   J.~C.~Collins and D.~E.~Soper,
%   %``Back-To-Back Jets: Fourier Transform from B to K-Transverse,''
%   Nucl.\ Phys.\ B {\bf 197}, 446 (1982).
%  
% \bibitem{Collins:1984kg} 
%   J.~C.~Collins, D.~E.~Soper and G.~F.~Sterman,
%   %``Transverse Momentum Distribution in Drell-Yan Pair and W and Z Boson 
%   %Production,''
%   Nucl.\ Phys.\ B {\bf 250}, 199 (1985). 



%------- replace following references ;-)
% \bibitem{parton_qed} A.~D.~Martin {\it et~al.} Eur. Phys. J. {\bf C39} (2005) 155.
% \bibitem{DVCS1}S.~Friot, B.~Pire and L.~Szymanowski. Phys. Lett. {\bf B645} (2007) 153.
% \bibitem{DVCS2}D.~Hasell, R.~Milner and K.~Takase. AIP Conf. Proc. {\bf 588} (2001) 187.
% \bibitem{DVCS3}M.~Krawczyk and A.~Zembrzuski. Phys. Rev. {\bf D57} (1998) 10.
% \bibitem{pomeron1}R.~Brower and C.~Tan. PoS LAT2005 (2006) 279.
% \bibitem{pomeron2}J.~P.~Guillaud and A.~Sobol,
%   ``Perspectives of the study of double Pomeron exchange at the LHC'', in
%   {\it 11th Lomonosov Conference on Elementary Particle Physics}, Moscow, Russia, 2003.
% \bibitem{Gogitidze:2007du} {\bfseries H1} Collaboration, N.~Gogitidze, ``Prompt photons and particle momentum distributions at hera''. hep-ex/0701033,
% 2007.
% \bibitem{H1}H1 Collab., N.~Gogitidze {\it et~al.}, ``{Prompt photons and particle
%   momentum distributions at HERA}'', 2007. 
% \href{http://arxiv.org/abs/hep-ex/0701033}{{\ttfamily hep-ex/0701033}}
\end{thebibliography}
\end{footnotesize}
}

% ****************************************************************************
% END OF BIBLIOGRAPHY AREA
% ****************************************************************************

\end{document}